\documentclass[doublecol]{epl2} 

\usepackage[T1]{fontenc}
\usepackage[latin9]{inputenc}
\usepackage{babel}
\usepackage{nccmath,amsmath,amssymb,graphics,xcolor}
\usepackage{subfig}

\title{Uncovering the root of LEFT in SMEFT}

\author{J. Chakrabortty\inst{1}\thanks{E-mail: \email{joydeep@iitk.ac.in}}
	 \and S. Prakash\inst{1}\thanks{E-mail: \email{surajprk@iitk.ac.in}} \and S. U. Rahaman\inst{1}\thanks{E-mail: \email{shakel@iitk.ac.in}} \and M. Spannowsky\inst{2}\thanks{E-mail: \email{michael.spannowsky@durham.ac.uk}}}
\shortauthor{J. Chakrabortty \etal}

\institute{                    
  \inst{1} Indian Institute of Technology Kanpur, Kalyanpur, Kanpur 208016, Uttar Pradesh, INDIA\\
  \inst{2} Institute for Particle Physics Phenomenology, Department of Physics, Durham University, Durham DH1 3LE, U.K.}

\abstract{
At energies below the electroweak scale, baryon number $B$ and lepton number $L$ violating processes are of significant importance in identifying the nature of UV extensions of the Standard Model. The imprint of UV theories on low-energy measurements can be calculated to a high accuracy using the theoretical framework of the Low Energy Effective Theory (LEFT). Using $B,\;L$ and the operators' dimensions as classifying characteristics, we construct a network connecting operator classes of the LEFT with the Standard Model Effective Theory (SMEFT). Following the links of this network, the contact interactions described by the effective operators of LEFT can be unambiguously embedded into those of SMEFT, which enables us to constrain the possible realisations of UV theories.  In turn, this can help to prioritise low energy measurements with the aim of comprehensively testing all classes of LEFT and SMEFT operators.}

\begin{document}

\maketitle

\section{Introduction}

Effective field theories (EFT) are established frameworks to provide a quantitative description of the impact of known or hypothesised particles with masses above the characteristic energy scale $E$ of a measurement. 
By integrating out degrees of freedom with masses heavier than the physical scale $\Lambda \gg E$, assuming a weakly coupled UV theory, EFTs correspond to a systematic expansion of the effective action in terms of local operators of dimension-$n$, suppressed by factors $\mathcal{O}(1/\Lambda^{n-4})$. As EFTs are a means of communication between physical models that reside at widely separated energy scales, they guide and shape our understanding in all areas of physics, from cosmology over nuclear or atomic physics to particle and high-energy physics. 

One of the phenomenologically most important theories is the Low Energy Effective Field Theory (LEFT) \cite{Jenkins:2017jig,Dekens:2019ept,Liao:2020zyx} of the Standard Model, which contains Fermi's theory of the effective interactions between four fermions and allows to describe the decay of a muon $\mu \to e \bar{\nu_e} \nu_\mu$ quantitatively to high accuracy. It arises after integrating out all particles with masses heavier than the bottom quark mass and is the EFT most suitable to describe weak decays of leptons, B-mesons and Kaons, thereby providing the theoretical foundation for the interpretation of measurements at B factories \cite{ATLAS:2020acx,LHCb:2020khb,LHCb:2020pcv}, neutrino experiments \cite{Mohapatra:2005wg,Arguelles:2019xgp,DUNE:2016hlj} and various beam-dump experiments \cite{Izaguirre:2013uxa,MiniBooNEDM:2018cxm}. 

To be precise, LEFT is defined with three generations of leptons and down type quarks, and two generations of up-type quarks and the underlying gauge symmetry of $SU(3)_C\otimes U(1)_{\text{em}}$. Low-energy observables, such as rare decays of $B$-mesons and Kaons are often very precisely measured, thereby exhibiting sensitivity to higher-dimensional operators in the power expansion of LEFT. Further, some well-measured processes are only affected by operators of dimension-9 and beyond within the LEFT. For both reasons, it is phenomenologically of fundamental importance to consider all LEFT operators to at least dimension-9. 

LEFT will cease to be a good description of nature when processes with energies $E\gtrsim m_W$ are considered. The encompassing theory to replace LEFT, viable to at least $\mathcal{O}(1)$ TeV, is the so-called Standard Model Effective Theory (SMEFT), which contains all effective interactions that can be induced by the particle content and gauge symmetries of the Standard Model, i.e. the gauge group $SU(3)_C\otimes SU(2)_L\otimes U(1)_{Y}$. Searches for the energy scale where SMEFT will have to be replaced by a more complete theory, and which theory takes this role, is currently the focus of the research programme at high-energy colliders, i.e. the LHC.

Baryon $B$ and lepton $L$ numbers are accidental global symmetries of the renormalisable Lagrangian of the Standard Model. They are expected to play a pivotal role in the explanation of the observed matter-antimatter asymmetry of the Universe \cite{Kuzmin:1985mm, Cohen:1993nk} and the finite masses of neutrinos \cite{Mohapatra:2005wg}. Interactions violating $B$ and $L$ would give rise to a plethora of phenomenologically striking processes, e.g. proton decay \cite{Lazarides:1980nt}, nucleon-nucleon oscillations \cite{Oosterhof:2019dlo}, rare decays of nucleons to three charged leptons \cite{Hambye:2017qix} and neutrinoless quadruple decays \cite{Heeck:2013rpa,Fonseca:2018aav}. All these rare processes would occur at low energy scales, well within the remit of LEFT, and have attracted attention as tangible evidence for new physics.

In this article, we report on the emergence of an intriguingly simple relation that connects the effective operator classes of LEFT and SMEFT when using the operators' dimension and the baryon and lepton quantum numbers as their classifying characteristics.  The unfolding of this network between operators does allow us to obtain the root of the LEFT operators in terms of the UV processes that underly the SMEFT framework. It will further allow us to obtain an improved understanding of the UV physics that sources the SMEFT operators.

\section{Classification of LEFT and SMEFT}\label{sec:leftsmeft}

LEFT contains spin-1/2 and spin-1 particles as its degrees of freedom. Working within $3+1$ dimensional space-time, where the Lorentz symmetry can be realised as $SU(2)\times SU(2)$ symmetry group, these degrees of freedom transform as  irreducible representations $\psi_L \equiv \left(\frac{1}{2}, 0\right)_{\frac{3}{2}}$, $\psi_R \equiv \left(0, \frac{1}{2}\right)_{\frac{3}{2}}$, $\mathfrak{X}_L \equiv \left(1, 0\right)_{2}$, $\mathfrak{X}_R \equiv \left(0, 1\right)_{2}$.
Here, $\psi_{L,R}$ are left and right chiral spinors and $\mathfrak{X}_{L,R}$ are the complexifications of the field strength tensor $\mathfrak{X}_{\mu\nu}$ and its dual $\widetilde{\mathfrak{X}}_{\mu\nu}$, expressed in two-component notation. The subscripts ($\frac{3}{2}$, $\frac{3}{2}$, $2$, and $2$) denote the mass dimension for each type of field. In addition to these fields we also include the covariant derivative $\mathsf{D}$ as a constituent of the invariant operators. It transforms as $\left(\frac{1}{2},\frac{1}{2}\right)$ representation of $SU(2)\times SU(2)$ and has unit mass dimension. 

With these constituents the construction of Lorentz invariant operators $\mathcal{O}$ can be schematically written as

{\small\begin{eqnarray}\label{eq:op-class-determination-1}
		\mathcal{O} &\equiv& \psi_L^{q_1}  \times \psi_R^{q_2} \times \mathsf{D}^{r} \times \mathfrak{X}_L^{s_1}  \times \mathfrak{X}_R^{s_2} \; ,
\end{eqnarray}}
\noindent
where $q_1, q_2, r, s_1, s_2$ are the number-of-times the different fields appear in the operator respectively. All these variables are non-negative integers. This relation can equivalently be written in terms of the fields' mass dimensions

{\small\begin{eqnarray}\label{eq:mass-dim}
		d &=& \frac{3}{2}(q_1+q_2) + r + 2(s_1 + s_2) \; ,
\end{eqnarray}}
where $d$ corresponds to the mass dimension of the Lorentz invariant operator.
Solving Eqn.~\eqref{eq:mass-dim} and  subsequently employing Eqn.~\eqref{eq:op-class-determination-1}, one can construct the Lorentz invariant operator classes for any given operator dimension. 
The dynamical nature of the fields and the presence of $\mathsf{D}$ relate some of the operator classes through equations of motion (EOMs) and integration by parts (IBPs). By exploiting these relations, we ensure to use non-redundant operator bases for our analysis. Apart from the space-time symmetry, internal symmetries, e.g., $SU(3)_C\otimes U(1)_{\text{em}}$ for LEFT, also play an important role in the formation of allowed operator classes.

For the discussion at hand the global symmetries $B,L$ are of profound importance \cite{Kobach:2016ami,Helset:2019eyc}. We can get an idea of the possible combinations of ($\Delta B$, $\Delta L$) at a given mass dimension in the following way. If the units for baryon and lepton numbers are $b$ and $l$ respectively and we have $n_1$ fermions carrying only baryon number and $n_2$ fermions carrying only lepton number\footnote{Fermions are usually assigned either baryon or lepton number, not both.}. Then the range of values of ($\Delta B$, $\Delta L$) is fixed by the requirements that $n_1+n_2 = q_1 + q_2$ and $n_1, n_2 \leq \frac{2}{3}d$. The equality holds only when either $n_1$ or $n_2$ is zero and $d \in 3\mathbb{N}$. Then, the range of ($\Delta B$, $\Delta L$) combinations are

{\small\begin{eqnarray}
		\bigg(|\Delta B|_{\text{min}},\, |\Delta L|_{\text{max}}\bigg) &=& \left(0,\,2\times\bigg\lfloor\frac{d}{3}\bigg\rfloor\times\big|l\big| \right),\nonumber\\ \hspace{0.5cm} \bigg(|\Delta B|_{\text{max}},\, |\Delta L|_{\text{min}}\bigg) &=& \left(2\times\bigg\lfloor\frac{d}{3}\bigg\rfloor\times\big|b\big|,\,0\right).
\end{eqnarray}}
Here, $\big\lfloor\cdot\big\rfloor$ denotes the floor function. Further restrictions are imposed based on the transformation properties of the fields under the internal symmetry.

We assign the $B, L$ charges to the degrees of freedom within $\text{LEFT}$ according to\footnote{In the literature different conventions for the assignment of $B$ and $L$ numbers are chosen. For the operator classes in LEFT and SMEFT we choose a consistent definition for the $(\Delta B, \Delta L)$ classification.}:  $u^s_{L,R},\, d^p_{L,R}\, \rightarrow\, (B = 1/3, \,L = 0);$ $\nu^p_L,\, e^p_{L,R} \rightarrow\, (B = 0, \,L = -1)$. Here, $p = 1,2,3$, whereas $s = 1,2$. This underlines the fact that the top-quark is already integrated out but all other fermions exhibit three flavours. Up to mass dimension-9, this results in the independent operator classes of LEFT
{\small\begin{eqnarray}\label{eq:LEFT-op-class}
		&\text{\textbf{\textit{d5}}}:& \psi^2\,\mathfrak{X}_{(0,0)}, \, \psi^2\,\mathfrak{X}_{(0,\text{-}2)} \;; \nonumber\\
		&\text{\textbf{\textit{d6}}}:& \mathfrak{X}^3_{(0,0)}, \, \psi^4_{(0,0)}, \, \psi^4_{(0,\text{-}2)}, \, \psi^4_{(1,\text{-}1)}, \,\psi^4_{(1,1)}, \,\psi^4_{(0,-4)}, \,\nonumber\\
		&&\underline{\mathsf{D}^2\mathfrak{X}^2_{(0,0)}}, \, \underline{\psi^2\mathfrak{X}\mathsf{D}_{(0,0)}} \;; \nonumber\\
		&\text{\textbf{\textit{d7}}}:&  \psi^4\,\mathsf{D}_{(0,0)}, \, \psi^4\,\mathsf{D}_{(0,\text{-}2)}, \, \psi^4\,\mathsf{D}_{(1,\text{-}1)}, \, \psi^4\,\mathsf{D}_{(1,1)},   \nonumber\\
		&& \psi^2\,\mathfrak{X}^2_{(0,0)}, \, \psi^2\,\mathfrak{X}^2_{(0,\text{-}2)}, \, \underline{\psi^2\mathfrak{X}\mathsf{D}^2_{(0,0)}}, \, \underline{\psi^2\mathfrak{X}\mathsf{D}^2_{(0,\text{-}2)}}   \;; \\
		&\text{\textbf{\textit{d8}}}:&  \mathfrak{X}^4_{(0,0)}, \, \psi^4\,\mathfrak{X}_{(0,0)}, \, \psi^4\,\mathfrak{X}_{(0,\text{-}2)}, \, \psi^4\,\mathfrak{X}_{(1,\text{-}1)}, \,   \psi^4\,\mathfrak{X}_{(1,1)},
		\nonumber\\
		&& \psi^4\,\mathfrak{X}_{(0,\text{-}4)}, \, \psi^4\,\mathsf{D}^2_{(0,0)}, \, \psi^4\,\mathsf{D}^2_{(0,\text{-}2)}, \, \psi^4\,\mathsf{D}^2_{(1,\text{-}1)}, \, \psi^4\,\mathsf{D}^2_{(1,1)}, \nonumber\\
		&& \psi^4\,\mathsf{D}^2_{(0,\text{-}4)}, \, \underline{\psi^2\,\mathfrak{X}^2\,\mathsf{D}_{(0,0)}}, \, \underline{\psi^2\mathfrak{X}\mathsf{D}^3_{(0,0)}}, \, \underline{\mathsf{D}^2\mathfrak{X}^3_{(0,0)}} \;;  \nonumber\\
		&\text{\textbf{\textit{d9}}}:& \psi^6_{(0,0)}, \, \psi^6_{(0,-2)}, \,  \psi^6_{(0,-4)}, \, \psi^6_{(0,-6)}, \, \psi^6_{(1,1)},  \, \psi^6_{(1,3)}, \nonumber\\
		&&  \psi^6_{(1,-1)}, \, 
		\psi^6_{(1,-3)}, \, \psi^6_{(2,0)}, \, \psi^4\,\mathfrak{X}\,\mathsf{D}_{(0,0)}, \, \psi^4\,\mathfrak{X}\,\mathsf{D}_{(0,-2)}, \nonumber\\
		&& \psi^4\,\mathfrak{X}\,\mathsf{D}_{(1,1)}, \, \psi^4\,\mathfrak{X}\,\mathsf{D}_{(1,-1)}, \,
		\psi^4\,\mathsf{D}^3_{(0,0)}, \, \psi^4\,\mathsf{D}^3_{(0,-2)}, \,   \psi^4\,\mathsf{D}^3_{(1,1)}, \nonumber\\
		&& \psi^4\,\mathsf{D}^3_{(1,-1)}, \,   \psi^2\,\mathfrak{X}^3_{(0,0)}, \, \psi^2\,\mathfrak{X}^3_{(0,\text{-}2)}, \, \underline{\psi^2\,\mathfrak{X}^2\,\mathsf{D}^2_{(0,0)}},  \nonumber \\
		&& \underline{\psi^2\,\mathfrak{X}^2\,\mathsf{D}^2_{(0,\text{-}2)}}, \, \underline{\mathsf{D}^3\mathfrak{X}^3_{(0,0)}}. \nonumber
	\end{eqnarray}}
Here, $\psi$, $\mathfrak{X}$ and $\mathsf{D}$ denote the fermion, field strength tensor, and covariant derivative for LEFT. This classification provides us a better handle to identify the relevant operator classes within the SMEFT.  The underlined operator classes can be related to other operator classes through appropriate equations of motion of the fields, hence these are excluded from the operator basis. For explicit details on these relations between operator classes, see \cite{Banerjee:2020jun}. We have also omitted operator classes that can be expressed as total derivatives, e.g. $\psi^2\mathsf{D}^2$ at dimension-5 or $\psi^2\mathfrak{X}\mathsf{D}^4$ at dimension-9. Since our analysis is aimed at finding correspondences of the operator classes of LEFT and SMEFT at the level of contact interactions, we have first identified the operator classes of SMEFT that have the same constitution as the LEFT ones or differ with respect to the presence of one or more scalars. We have explicitly demarcated these classes based on baryon and lepton number violation. After noting down the Lorentz invariant operator classes for SMEFT, we have employed the EOM and IBP relations to identify the independent ones, see \cite{Banerjee:2020jun} for a pedagogical account. Then we assign the $B$ and $L$ quantum numbers to the SM degrees of freedom as:
$Q^p_L,\, u^p_R,\, d^p_R\, \rightarrow\, (B = 1/3, \,L = 0)$ $L^p_L,\, e^p_R \rightarrow\, (B = 0, \,L = -1)$ to obtain the operator classes relevant for this analysis. Here, $p = 1,2,3$.

{\small\begin{eqnarray}\label{eq:SMEFT-op-class}
		&\hspace*{-0.5cm}\text{\textbf{\textit{D6}}}:& X^3_{(0,0)}, \, \Psi^4_{(0,0)}, \, \Psi^4_{(1,\text{-}1)}, \, \Psi^2\Phi X_{(0,0)}; \nonumber\\
		&\hspace*{-0.5cm}\text{\textbf{\textit{D7}}}:& \Psi^4\mathcal{D}_{(0,\text{-}2)}, \, \Psi^4\mathcal{D}_{(1,1)}, \, \Psi^4\Phi_{(0,\text{-}2)}, \, \Psi^4\Phi_{(1,1)}, \,\Psi^2\Phi^2X_{(0,\text{-}2)}; \nonumber\\
		&\hspace*{-0.5cm}\text{\textbf{\textit{D8}}}:&  X^4_{(0,0)}, \, \Psi^4\,X_{(0,0)}, \, \Psi^4X_{(1,\text{-}1)}, \,\Psi^4\mathcal{D}^2_{(0,0)}, \, \Psi^4\mathcal{D}^2_{(1,\text{-}1)},  \nonumber\\
		&\hspace*{-0.5cm}& \Psi^2X^2\mathcal{D}_{(0,0)}, \, \Psi^4\Phi\mathcal{D}_{(0,0)}, \, \Psi^4\Phi\mathcal{D}_{(1,\text{-}1)}, \, \Psi^2\Phi X^2_{(0,0)}; \nonumber\\
		&\hspace*{-0.5cm}\text{\textbf{\textit{D9}}}:&\Psi^2\Phi^2X^2_{(0,\text{-}2)}, \,  \Psi^4\Phi X_{(1,1)}, \, \Psi^4\Phi X_{(0,\text{-}2)}, \, \Psi^4\Phi\mathcal{D}^2_{(1,1)}, \nonumber\\
		&\hspace*{-0.5cm}& \Psi^4\Phi\mathcal{D}^2_{(0,\text{-}2)}, \, \Psi^4X\mathcal{D}_{(0,\text{-}2)}, \, \Psi^4X\mathcal{D}_{(1,1)} \, \Psi^4\mathcal{D}^3_{(0,\text{-}2)}, \nonumber\\
		&\hspace*{-0.5cm}& \Psi^4\mathcal{D}^3_{(1,1)}, \, \Psi^6_{(0,\text{-}2)}, \, \Psi^6_{(1,1)}, \,  \Psi^6_{(1,\text{-}3)}, \, \Psi^6_{(2,0)};  \nonumber\\
		&\hspace*{-0.5cm}\text{\textbf{\textit{D10}}}:& \Psi^2\Phi X^3_{(0,0)}, \,  \Psi^4\Phi X \mathcal{D}_{(0,0)}, \, \Psi^4\Phi X \mathcal{D}_{(1,\text{-}1)}, \, \Psi^4\Phi\mathcal{D}^3_{(0,0)}, \nonumber \\
		&\hspace*{-0.5cm}& \Psi^4\Phi\mathcal{D}^3_{(1,\text{-}1)}, \, \Psi^4\,\Phi^4_{(0,\text{-}4)}, \, \Psi^6\Phi_{(0,0)} \, \Psi^6\Phi_{(1,\text{-}1)}  \, \Psi^6\Phi_{(1,\text{-}3)}; \nonumber\\
		&\hspace*{-0.5cm}\text{\textbf{\textit{D11}}}:& \Psi^2\Phi^2X^3_{(0,\text{-}2)}, \, \Psi^6\Phi^2_{(0,\text{-}4)}; \\
		&\hspace*{-0.5cm}\text{\textbf{\textit{D12}}}:& \Psi^4\Phi^4\,X_{(0,\text{-}4)}, \, \Psi^4\Phi^4\mathcal{D}^2_{(0,\text{-}4)}; \nonumber\\
		&\hspace*{-0.5cm}\text{\textbf{\textit{D15}}}:& \Psi^6\Phi^6_{(0,\text{-}6)}. \nonumber
\end{eqnarray}}
Here, $\Psi$, $\Phi$, $X$ and $\mathcal{D}$ denote the fermion, scalar, field strength tensor, and covariant derivative for SMEFT. $(\Delta B, \Delta L)$ are explicitly provided for each operator class as subscripts in Eqns.~\ref{eq:LEFT-op-class} and \ref{eq:SMEFT-op-class}. Note that the SMEFT operator classes listed above are the lowest mass dimensional classes that encapsulate the LEFT classes with the same ($\Delta B, \Delta L$) values. But if one investigates individual operators as opposed to the schematic classes, one may find the root of certain operators in even higher mass dimensional SMEFT operators. For instance, while the $\psi^6_{(0,\text{-}4)}$ class of LEFT (at $d9$) is embedded within the $\Psi^6\Phi^2_{(0,\text{-}4)}$ class of SMEFT (at $D11$), certain operators containing 4 neutrino fields find their origin at $\Psi^6\Phi^4_{(0,\text{-}4)}$ class at $D13$. The contents of Eqn.~\ref{eq:SMEFT-op-class} highlight the lowest mass dimension of SMEFT where our search for the roots of a particular operator class of LEFT must begin.

\section{Relating Wilson Coefficients of LEFT and SMEFT}

One of the  most notable differences between LEFT and SMEFT arises through the Higgs field ($\Phi$), which acquires a vacuum expectation value (vev), i.e. $\langle\Phi \rangle$, and induces the spontaneous breaking of the electroweak symmetry (EWSB). 

For matching at the level of contact interactions between operator classes of LEFT and SMEFT having the same $(\Delta B, \Delta L)$ values but differing with respect to the presence of one or more scalars, powers of the vacuum expectation value $\langle\Phi\rangle$  appear as scaling factors. For example, the Wilson Coefficients of the elements of an $n$-dimensional SMEFT operator class containing $k$ scalar fields can be matched onto the Wilson Coefficients of the elements of an $(n$-$k)$-dimensional LEFT operator class scaled by $\langle\Phi \rangle^k$.

The bosonic degrees of freedom in LEFT consist of the photon and gluons only, i.e. $\mathcal{V}_{\mu} = A_{\mu},\, G^a_{\mu}$\footnote{While $\mathfrak{X}$ ($\mathfrak{X}_{\mu\nu}$) and $X$ ($X_{\mu\nu}$) have been used to schematically denote the field strength tensors of LEFT and SMEFT respectively, the corresponding gauge bosons have been denoted by $\mathcal{V}_{\mu}$ and $V_{\mu}$.}, whereas for SMEFT we have $V_{\mu} = W^I_{\mu},\, B_{\mu}$, and $G^a_{\mu}$\footnote{Here, $I=1,2,3$ and $a=1,\cdots,8$ are the $SU(2)_L$ and $SU(3)_C$ gauge indices respectively.}. Hence, their operators consist of the corresponding field strength tensors $F_{\mu\nu},\, G^a_{\mu\nu}$ and $W^I_{\mu\nu},\, B_{\mu\nu},\, G^a_{\mu\nu}$ respectively. EWSB further leaves a significant impact by rotating the electroweak gauge bosons ($W^I_{\mu},\, B_{\mu}$) into a physical basis ($W^{\pm}_{\mu},Z_{\mu},A_{\mu}$).  An important step leading from SMEFT to LEFT is the  mixing between the electrically neutral bosonic degrees of freedom
{\small\begin{eqnarray}\label{eq:sm-weak-mixing}
		\begin{pmatrix}
			W^3_{\mu} \\ B_{\mu}
		\end{pmatrix} = \begin{pmatrix}
			\cos\theta_{\text{w}} \ \ \sin\theta_{\text{w}}\\ -\sin\theta_{\text{w}} \ \ \cos\theta_{\text{w}}
		\end{pmatrix} \begin{pmatrix}
			Z_{\mu} \\ A_{\mu}
		\end{pmatrix}.
\end{eqnarray}}
Thus at the operator-class level there are two explicit factors that play a crucial role in this matching:
\begin{enumerate}
	\item Replacing the Higgs doublet with its vev $\langle\Phi\rangle$.
	
	\item Rotating the gauge bosons from the unphysical $W^3_{\mu}, B_{\mu}$ to the physical $Z_{\mu}, A_{\mu}$ basis.  In the process we discard the heavy field tensors $W^{\pm}_{\mu \nu}, Z_{\mu \nu}$.  We employ the rotation in the gauge field sector, Eqn.~\ref{eq:sm-weak-mixing}, and record the matching coefficients. For instance, a generic operator containing $(B_{\mu\nu})^l\,(W^I_{\mu\nu})^m$, would lead to factors of $\cos^l{\theta_{\text{w}}}\,\sin^m{\theta_{\text{w}}}$. It is important to note that not all combinations of $l$ and $m$ would be relevant to the matching.
\end{enumerate}

\begin{figure}[!htb]
	\centering
	{
		\includegraphics[width=7.5cm, height=12cm, trim =20 10 0 10]{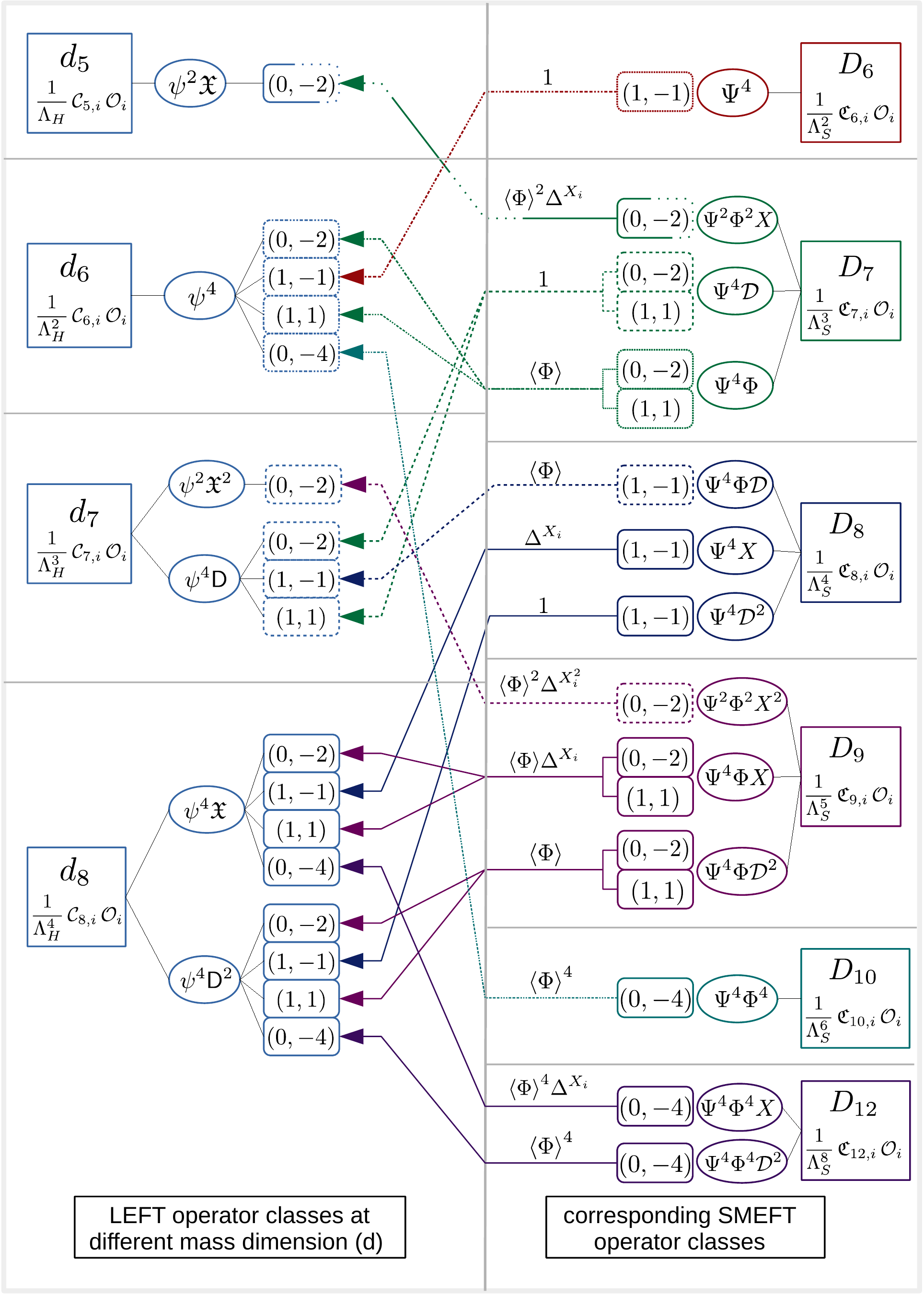}
	}
	\caption{\small The schematic embedding of contact interactions encapsulated within LEFT operator classes up to mass dimension-8 into the SMEFT contact interactions. Starting from a particular operator class of LEFT, with specific ($\Delta B,  \Delta L$) values, tracing backwards along the arrow one can arrive at the relevant SMEFT operator class. Here, $\Delta^{X_i}, \Delta^{X^2_i}$ refer to factors of $\cos\theta_{\text{w}}$ or $\sin\theta_{\text{w}}$ depending on the SMEFT operator being linear or quadratic in $B_{\mu\nu}$ or $W^I_{\mu\nu}$ and $\langle\Phi\rangle$ denotes the vacuum expectation value of the SM Higgs.  }
	\label{fig:LEFT-SMEFT}
\end{figure}

The results of this matching procedure between SMEFT and LEFT, and the interrelations arising from it have been depicted in full detail in Fig.~\ref{fig:LEFT-SMEFT}. We have also shown the multiplicative factors that appear for each case and the values of $\Delta B$ and $\Delta L$. One must be careful while performing the matching of individual operators, as opposed to the operator classes, since that may reveal certain subtle aspects pertaining to the relationship between LEFT and SMEFT operators. Some of the additional steps that need to be performed on a case-by-case basis while matching the operators of LEFT and SMEFT, see \cite{Murphy:2020cly,Li:2020gnx,Murphy:2020rsh,Liao:2020jmn,Li:2020xlh}:	

\begin{enumerate}
	
	\item The charged fermion mixing matrix, e.g.,  the Cabibbo-Kobayashi-Masakawa matrix in quark sector $V_{ij}$, where, $i$=$u,c,t$ and $j$=$d,s,b$.
	\item  The modified flavour structures of the LEFT operators as the $top-$quark must be integrated out from SM(EFT) leaving only two up and three down-type quarks.
	
	\item The impact of rotation matrix in neutral gauge boson sector in the inner structures of covariant derivatives $\mathsf{D}$, derived from  $\mathcal{D}$. 
\end{enumerate}

The detailed analysis is beyond the scope of the current discussion. Our aim, in this work, is to indicate those SMEFT classes which encapsulate contact interactions in which the LEFT classes with similar field content and identical $(\Delta B, \Delta L)$  violations can be shown to be embedded. This will be helpful to find the UV roots of rare processes induced by the LEFT operators through SMEFT ones.

\section{Lessons from the LEFT cut-off scale}

While discussing the interrelations between two effective theories it is pertinent to remark on the high energy scales that act as suppression factors for the operators. These can be entirely arbitrary or have well defined relations. For the case of LEFT and SMEFT (with $\Lambda_H$ and $\Lambda_S$ being the respective suppression scales), we identify two distinct scenarios:

\begin{enumerate}
	\item $\Lambda_{H} \simeq m_W$: Both the internal symmetry and field content of LEFT are subsets of those of the Standard Model. Thus, we can obtain certain LEFT operators by integrating out the heavy SM degrees of freedom, i.e. $W^{\pm}_{\mu}, \,Z_{\mu}, H, t$. We emphasise that the renormalisable SM Lagrangian $\mathcal{L}_{SM}^{ren}$ possesses accidental global symmetries, like  baryon $B$ and lepton $L$ numbers. Thus, only the $B,L$ conserving LEFT operators can originate from $\mathcal{L}_{SM}^{ren}$. 
	
	\item $\Lambda_H \in \{\Lambda_S, \Lambda_{_{BSM}}\}$: Following the previous argument, all the $\Delta B, \Delta L \neq 0$  LEFT operator classes must originate in beyond SM renormalisable  interactions that violate $B,L$ numbers explicitly. If the scale of new physics $\Lambda_{_{BSM}}$ is much larger than the EW scale, the leading contributions will appear through SMEFT-operators with $\Delta B, \Delta L \neq 0$. In that case $\Lambda_H=\Lambda_S$ modulo quantitative modifications due to the RGE running of the operators. However, in some UV scenarios new degrees of freedom may reside at energy scales $\Lambda_{_{BSM}} \lesssim 100$ GeV, which have not been discovered to date and are not part of the SMEFT framework. The existence of such particles can give rise to the classes of LEFT operators and modified hierarchies of scales  $\Lambda_H\simeq \Lambda_{_{BSM}} \ll \Lambda_S$. Such LEFT operators may not be very suppressed and can lead to some rare experimentally observable processes. Whether these processes are observed or not will provide a better understanding of the cut-off scale and the mass scale of the BSM particles. 
	
\end{enumerate}

Additionally, if we consider, for instance, a process described by a tree-level diagram involving a scalar propagator and where one vertex is a renormalizable SM vertex with $(\Delta B = 0, \,\Delta L = 0)$ (say $\Psi^2\Phi$) and the other is a non-renormalizable effective vertex  of dimension-7 (e.g., $\Psi^4\Phi$) that violates baryon and (or) lepton number(s) ($\Delta B = 0, \Delta L = \text{-}2$ in this case), then we expect to obtain the LEFT operator class $\psi^6_{(0,\text{-}2)}$, which we have matched with the SMEFT contact term described by $\Psi^6_{(0,\text{-}2)}$ at dimension-9. So, it may appear that the $\Psi^4\Phi_{(0,\text{-}2)}$ class gives the leading contribution to LEFT operator class. But, our entire analysis is aimed at finding SMEFT operator classes within which the LEFT classes are embedded as contact interactions. So, a SMEFT class at dimension-$n$ cannot embed a LEFT class of mass dimension $\geq n$, rather two SMEFT classes can be involved in a process that may generate the LEFT class after integrating out a heavy field but this is different from identifying the embedding of LEFT operator classes within SMEFT contact interactions, which is the aim of this work. Secondly, even if we take into account the integrating out procedure involving the SMEFT classes $\Psi^2\Phi$, $\Psi^4\Phi_{(0,\text{-}2)}$ and leading to the LEFT class $\psi^6_{(0,\text{-}2)}$ if the coupling constant ($y$) involved in the renormalizable vertex assumes a value such that $\frac{y}{m^2_H} \sim \frac{1}{\Lambda_{UV}^2}$ then the contribution of the contact interaction cannot be deemed subleading.

The BSM scenarios contain additional degrees of freedom relative to the SM. If we consider UV theories, like Pati-Salam \cite{Pati:1974yy}, Left-Right Symmetric Models \cite{Mohapatra:1974hk,Chang:1983fu}, Grand Unified Theories \cite{Gursey:1975ki}, they induce $B$ and(or) $L$ violating interactions \cite{Weinberg:1980bf,Klapdor-Kleingrothaus:2002rvk}. As a consequence, these theories predict Majorana neutrino mass terms, neutron-antineutron oscillation, proton decay,  charged lepton flavour and number violation, neutrinoless double and quadruple beta decays, which are striking signatures for $\Delta B, \Delta L \neq 0$ processes. These rare processes do not suffer from any SM background, making them outstanding channels for BSM searches.

Observation of such rare events will clearly signify the presence of new physics. This provides motivation to classify the LEFT operators that can give rise to these rare signatures. It is further important to know their root in SMEFT, such that it allows to trace the mass scale of new physics, i.e. UV theories. Hence, for the categorisation of operators, we are focusing on classes that mediate $\Delta B, \Delta L \neq 0$.

\section{Spectroscopy of the UV through operator networks}
\label{sec:spectroscopy}

LEFT operators only contain the lighter fermions as fundamental fields. As hadrons are bound states of these quark fields, LEFT provides the most direct theoretical framework to describe rare hadronic processes \cite{Hambye:2017qix,Oosterhof:2019dlo}.  

\begin{figure}[!htb]
	\centering
	{
		\includegraphics[width=8cm, height=10cm, trim =20 10 0 10]{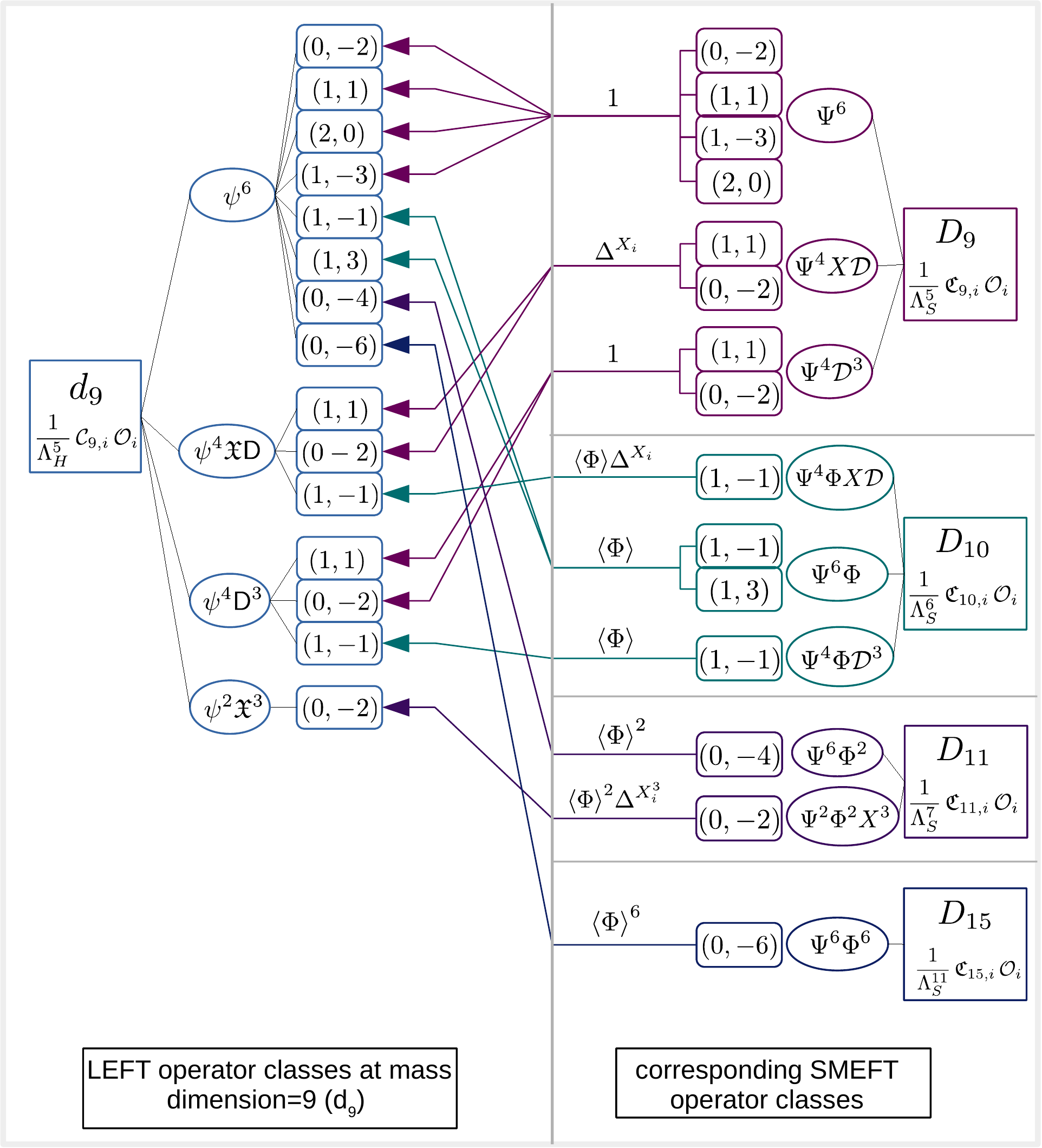}
	}
	\caption{\small The schematic embedding of contact interactions encapsulated within LEFT operator classes of mass dimension-9 into higher dimensional ($\geq 9$) SMEFT contact interactions. Once again, $\Delta^{X_i}, \Delta^{X^3_i}$ refer to factors of $\cos\theta_{\text{w}}$ and $\sin\theta_{\text{w}}$ and $\langle\Phi\rangle$ denotes the vacuum expectation value of the SM Higgs.}
	\label{fig:LEFT-SMEFT9}
\end{figure}

Based on the results so far, we can infer which UV theory can give rise to processes measured at energy scales of $\mathcal{O}(10)$ GeV or less. At such energy scales the interpretation framework suitable to describe nature to a high accuracy is LEFT.  
Interesting predictions can be made by studying contact interactions involving 6 fermions (with various combinations of quarks and leptons). Such phenomena are encountered starting at mass dimension-9. LEFT has 19 operator classes at mass dimension-9 (15 if we exclude the $B$-, $L$- conserving ones) as shown in Eqn.~\eqref{eq:LEFT-op-class} and these are embedded within SMEFT contact interactions that arise at mass dimension-9  up to dimension-15. We have shown this network in Fig.~\ref{fig:LEFT-SMEFT9}.

\begin{figure}[!htb]
	\renewcommand\thesubfigure{\roman{subfigure}}
	\subfloat[]{\includegraphics[width=8.75cm, height=2.8cm,trim =5 8 10 20]{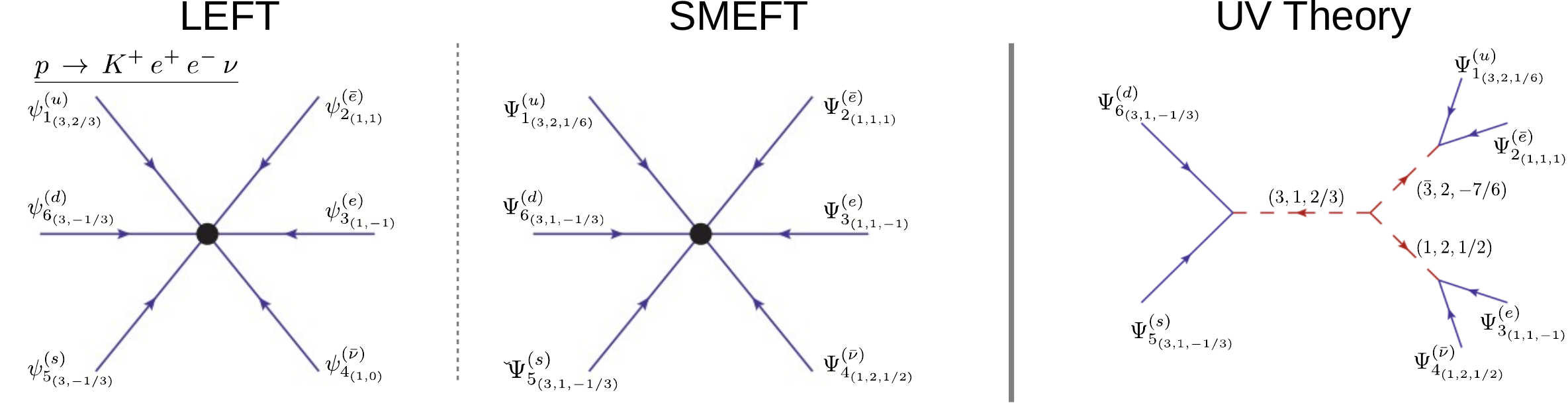}}
	\newline
	\subfloat[]{\includegraphics[width=8.75cm, height=2.8cm,trim =5 8 10 20]{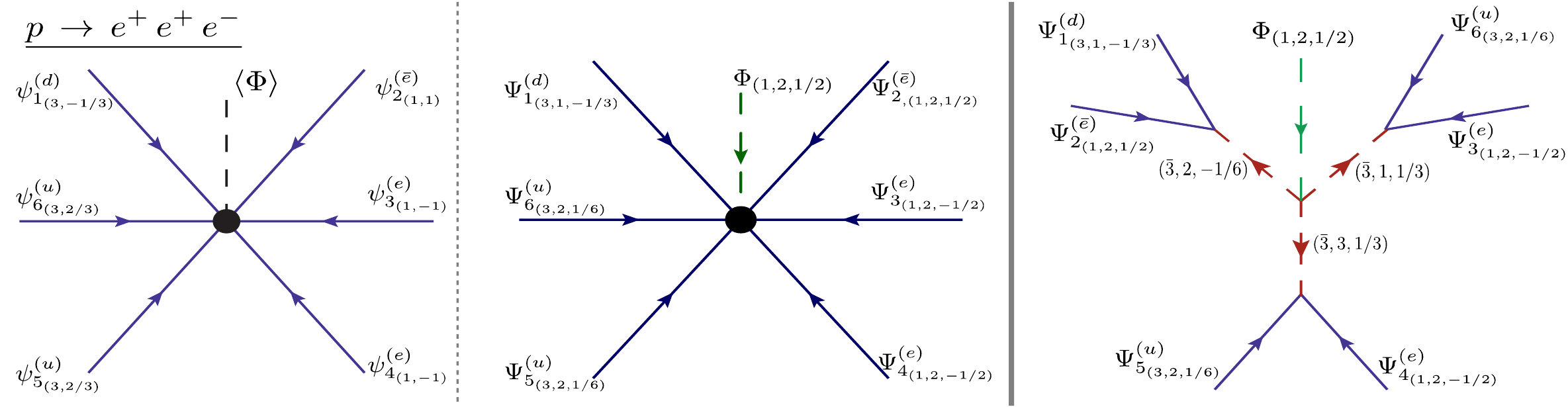}}
	\caption{\small Unfolding of LEFT operators that facilitate the rare processes: (i)  $p\to K^{+} e^{+} e^{-} \nu$ and (ii) $p\to e^{+} e^{+} e^{-}$, through the exchanges of BSM particles at tree level diagrams of UV theories. Here, solid and dashed lines correspond to fermions and bosons respectively. The red coloured lines denote the particles which are to be integrated out and the green ones denote the Higgs which will be replaced by its vev. The numbers within the parenthesis denote quantum numbers under the corresponding gauge groups. Within each diagram, the first column shows LEFT operartors, the second depicts the corresponding SMEFT operators and the last column contains the possible UV roots.}
	\label{fig:rare-processes}
\end{figure}

\begin{figure}[!htb]
	\subfloat[]{\includegraphics[width=2.6cm, height=5.9cm]{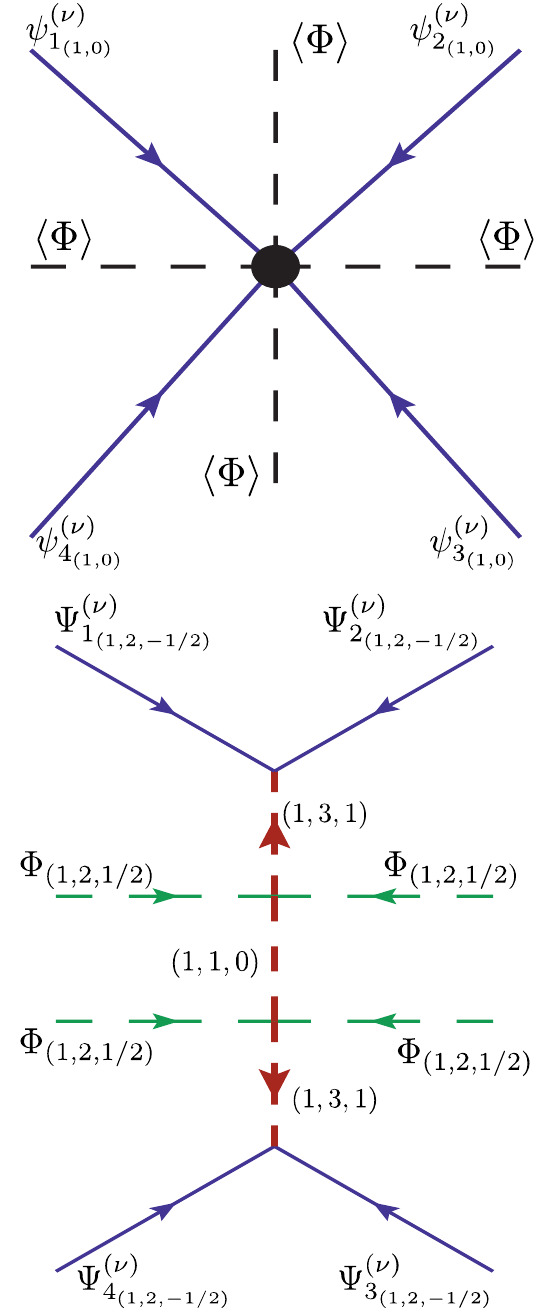}}
	\subfloat[]{\includegraphics[width=2.6cm, height=5.9cm]{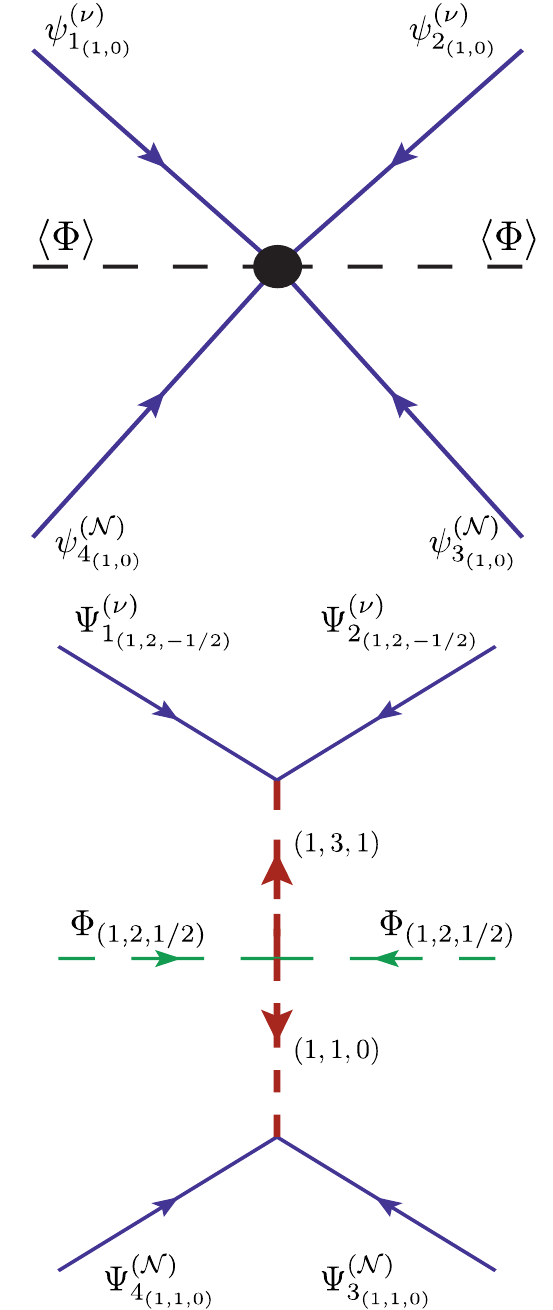}}
	\subfloat[]{\includegraphics[width=2.6cm, height=5.9cm]{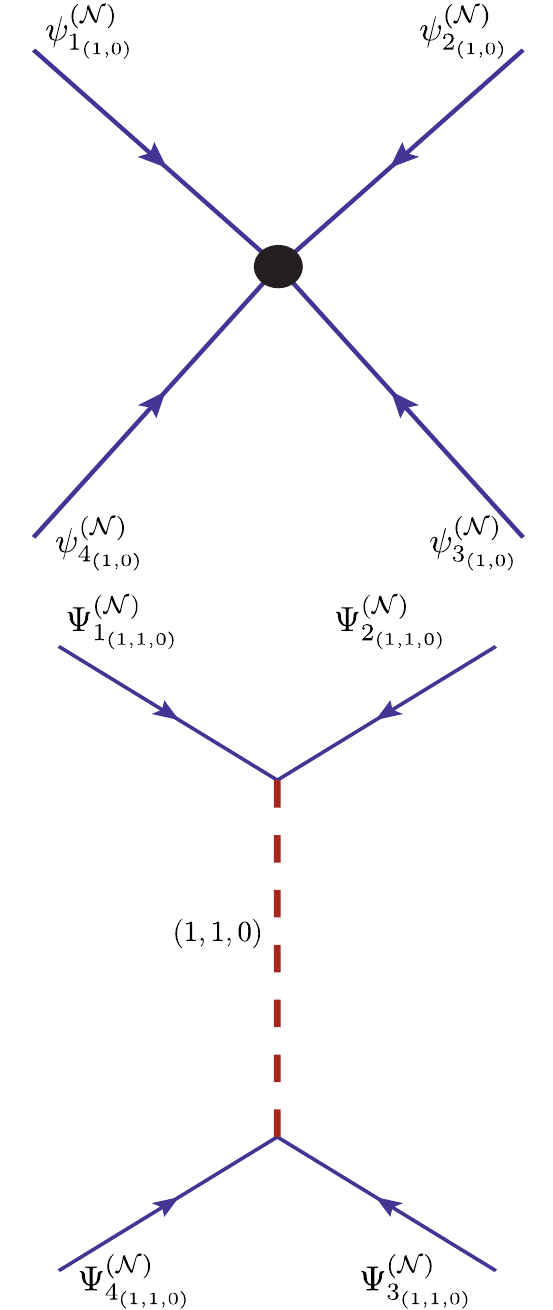}}
	\caption{\small Unfolding low energy operators for three distinct  cases: (a) $\Psi_i\in \nu$ (LEFT), (b)  $\Psi_i\in \{\nu, \mathcal{N}\}$, and (c)  $\Psi_i \in \mathcal{N}$ leading to (iii) $(A,Z)\to (A,Z+4)+4e^{-}$. The red colored lines denote the particles which are to be integrated out and the green ones denote the Higgs which will be replaced by its vev.}
	\label{fig:left-lneft-comp}
\end{figure}

The connections highlighted in Fig.~\ref{fig:LEFT-SMEFT9} are vital for explorations within the context of hadronic physics. To make the ideas concrete, we discuss processes that are prohibited by $B$ and $L$ conservation, e.g. (i) $p\to K^{+} e^{+} e^{-} \nu$, (ii) $p\to e^{+} e^{+} e^{-} $ shown in Fig.~\ref{fig:rare-processes}, and (iii) $(A,Z)\to (A,Z+4)+4e^{-}$ shown in Fig.~\ref{fig:left-lneft-comp}. An observation of any of these processes at existing experiments would signify new physics at a relatively low energy scale. The processes (i) and (ii) arise at dimension-9 in LEFT through the operators $\psi^6_{(1,1)}$ and $\psi^6_{(1,-1)}$ respectively, whereas process (iii) arises already at dimension-6 through $\psi^4_{(0,-4)}$.

After identifying the higher mass dimensional contact interactions of LEFT that give rise to the physical processes (i)-(iii), one can directly pin down the operator classes within SMEFT containing the relevant contact interactions using Fig.~\ref{fig:LEFT-SMEFT9}. In a second step, the SMEFT operators can be unfolded into tree-level diagrams by integrating in the fundamental degrees of freedom of a UV theory \cite{DasBakshi:2018vni,Banerjee:2020bym}. We show these steps explicitly from left to right in Fig.~\ref{fig:rare-processes} for processes (i) and (ii) and in Fig.~\ref{fig:left-lneft-comp} for process (iii).  

An interesting situation arises when the LEFT operator does not connect external particles of a physical process. For example, in the case of process (iv), mediated through the operator class $\psi^4_{(0,-4)}$, where $\psi$'s are the light active neutrinos $\nu$, this operator is embedded in $D_{10}$, see Fig.~\ref{fig:LEFT-SMEFT}, which can be expanded by inserting heavy BSM propagators, as shown in Fig.~\ref{fig:left-lneft-comp}(a). 

Our construction, i.e. the network of Figs.~\ref{fig:LEFT-SMEFT} and \ref{fig:LEFT-SMEFT9}, is equally applicable if light non-SM particles exist that mimic the LEFT degrees of freedom \cite{Banerjee:2020jun,Banerjee:2020bym}. Their presence will not alter the LEFT operator classes, rather offer more operators within the same class. For example,  a very light sterile neutrino $\mathcal{N}$ possesses identical $SU(3)_C\otimes U(1)_{\text{em}}$ quantum number as $\nu$. Thus, the $\psi^4_{(0,-4)}$ operator class with $\Psi_i\in \{\nu, \mathcal{N}\}$ is augmented by operators that contribute to the same process, see Figs.~\ref{fig:left-lneft-comp} (b) and (c). However, as there is a variety of UV theories that can give rise to $\nu$ and $\mathcal{N}$ simultaneously, there are several possibilities to mediate these processes, see Figs.~\ref{fig:left-lneft-comp} (a), (b), and (c). These operators carry a different mass dimension and the experimental sensitivity to them is therefore vastly different. Hence, the LEFT operator classification and its associated network are not restricted to scenarios where the lighter degrees of freedom are part of the SM particle content only. This classification also qualifies to capture the presence of LEFT look alike BSM IR degrees of freedom and  their UV roots can also be traced following the same strategy. 

Thus, the network of Figs.~\ref{fig:LEFT-SMEFT} and \ref{fig:LEFT-SMEFT9} unveils the embedding of LEFT operator classes within SMEFT contact interactions and their further UV expansion leaves us with a very restricted set of possibilities of BSM particles and theories.  

\section{Conclusion}
\label{sec:conclusion}

At low energies, far below the electroweak scale, experiments are searching for rare processes as signatures for physics beyond the Standard Model. Many of these processes are induced by an explicit violation of $B$ and(or) $L$ numbers, accidental global symmetries at the perturbative level within the Standard Model. Thus they are not expected to be generated through interactions of the renormalisable SM Lagrangian. Observing such processes at current or future collider experiments would indicate the realisation of new physics at accessible energy scales. The theoretical framework to describe such processes at flavour factories, beam dump or neutrino experiments is LEFT, the low energy effective theory. We have provided the phenomenologically important classes of LEFT operators up to dimension-9, with a focus on $B$ and $L$ violating operators. By identifying how the LEFT operators descend from the contact interactions of the  SMEFT framework, or its extensions, we can localise the energy scale of new physics and constrain the nature of UV theories that give rise to rare processes, observable at low energy experiments. Conversely, it allows us to prioritise experimental searches that are sensitive to physics within an accessible energy range.

\acknowledgments
The work of JC, SP, SUR is supported by the Science and Engineering Research Board,
Government of India, under the agreements SERB/PHY/2016348 (Early Career Research
Award) and SERB/PHY/2019501 (MATRICS) and Initiation Research Grant, agreement
number IITK/PHY/2015077, by IIT Kanpur. M.S. is supported by the STFC under grant
ST/P001246/1.

\end{document}